\documentclass{article}
\usepackage{spconf,amsmath,graphicx}
\usepackage{mwe} 
\usepackage{amssymb}
\usepackage{amsfonts}
\usepackage{booktabs}

\def\pos{{\mathbf x}}
\title{Combining Bayesian And Deep Learning Methods For The Delineation Of The Fan In Ultrasound Images}

\name{Hind Dadoun$^{\star}$ \quad Hervé Delingette$^{\star}$ \quad Anne-Laure Rousseau$^{\dagger}$ \quad Eric de Kerviler$^{\ddagger}$ \quad Nicholas Ayache$^{\star}$}

\address{$^{\star}$ Université Côte d'Azur, Inria Epione Team, Sophia Antipolis, France \\
    $^{\dagger}$ Hôpital Européen Georges Pompidou, NHance, Nabla, Paris, France \\
     $^{\ddagger}$ Radiology, Saint Louis Hospital, AP-HP, Paris, France \\
    }
\usepackage[belowskip=-15pt,aboveskip=0pt]{caption}
\begin{document}
%
\maketitle
\begin{abstract}
  Ultrasound (US) images usually contain identifying information outside the ultrasound fan area and manual annotations placed by the sonographers during exams. For those images to be exploitable in a Deep Learning framework, one needs to first delineate the border of the fan which delimits the ultrasound fan area and then remove other annotations inside. We propose a parametric probabilistic approach for the first task. We make use of this method to generate a training data set with segmentation masks of the region of interest (ROI) and train a U-Net to perform the same task in a supervised way, thus considerably reducing computational time of the method, one hundred and sixty times faster. These images are then processed with existing inpainting methods to remove annotations present inside the fan area. To the best of our knowledge, this is the first parametric approach to quickly detect the fan in an ultrasound image without any other information than the image itself.

\end{abstract}
\section{Introduction}
\label{sec:intro}

Ultrasound (US) imaging is one of the most common techniques for medical diagnosis. According to the WHO, two thirds of the world’s population do not have access to medical imaging, and ultrasound associated with X-ray could cover 90 \% of these needs.  The decrease in ultrasound hardware prices allows its diffusion, but limitations persist because acquiring and interpreting an ultrasound image is a difficult examiner-dependent task with few trained operators\cite{moore2011point,liebo2011pocket, choi2011interpretation}.
Hence the importance of developing the research around the entire processing chain in US imaging \cite{droste2020automatic,van2019automated,yap2017automated}. 
Interested readers may refer to \cite{van2019deep} for an overview of Deep Learning strategies for ultrasound-specific processing methods. 

\begin{figure}[htb!]
\begin{minipage}[b]{0.48\linewidth}
  \centering
  \centerline{\includegraphics[width=4.5cm]{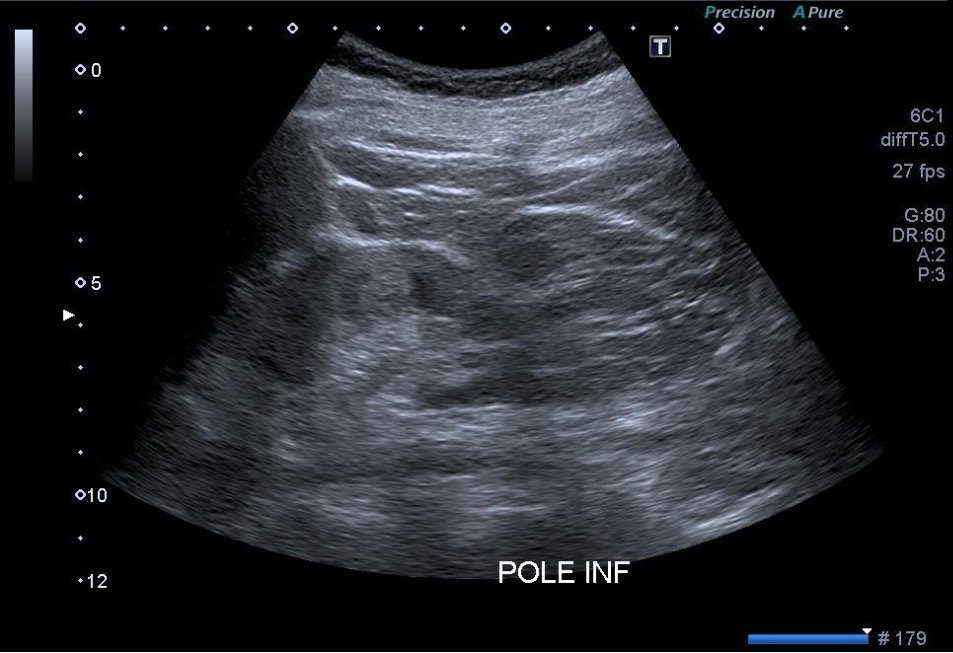}}
\end{minipage}
\begin{minipage}[b]{0.48\linewidth}
  \centering
  \centerline{\includegraphics[width=4.48cm]{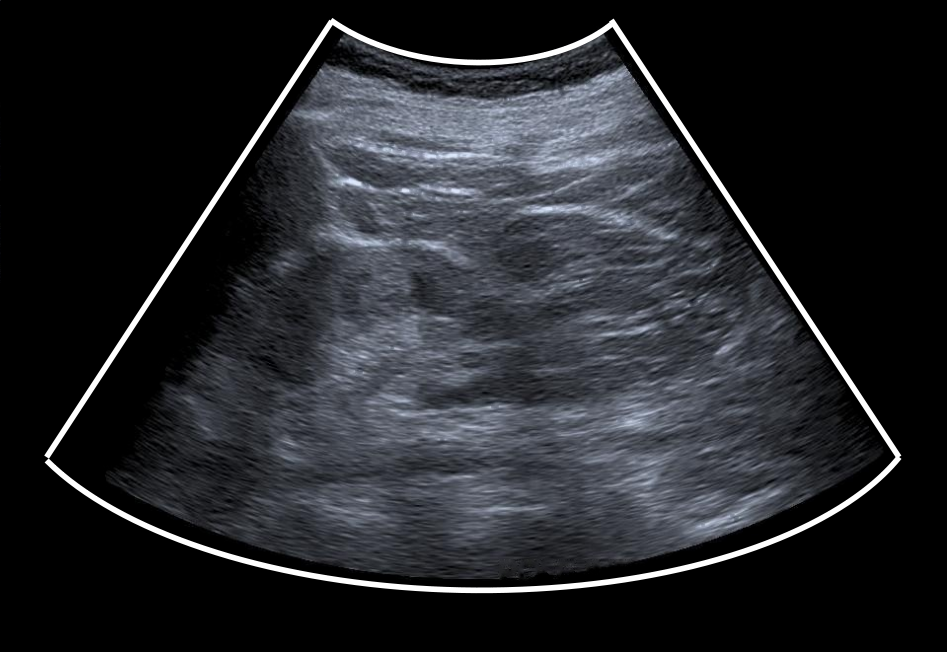}}
\end{minipage}
\vspace{0.6cm}
\caption{Left picture shows the original image. One can see that the ultrasound fan area is limited to a conic section, and that a number of text and graphic elements are present. Right picture shows the result of our pre-processing.The white lines delimiting the ultrasound fan area are automatically detected and all graphic and text elements are removed and replaced by a plausible intensity value. }
\label{fig:final}
\end{figure}
Ultrasound imaging provides real-time anatomical information and allows the measurement of functional parameters. These measurements along with other annotations are manually added to the image during an exam to provide a complete report. 
In order to make them available to the  research community following regulatory guidelines, they are usually converted to  JPEG/PNG format, and processed to remove all metadata, including acquisition parameters. The presence of biometric measurements and other machine dependent characteristics may be challenging for the task of automated image analysis. First, because there is no guarantee that these annotations do not include sensitive information. Second, because we would like to make sure that they do not induce a bias during the training of a neural network for the task of classification, segmentation or detection.

In \cite{zhang2017computer}, the authors “blacked  out”  the  identifying patient information on videos by setting the corresponding intensities of pixels that remained static throughout the entire clip to minimal intensity. Unfortunately, this method only works if we have access to the video sequence of the exam.  As for the biometric measurements, \cite{baumgartner2017sononet} removed all the annotations using the inpainting algorithm proposed in \cite{telea2004image}, but no information was given on how to create the inpainting masks. In this work we propose a pre-processing pipeline for ultrasound images to detect the ultrasound fan area combining Bayesian and Deep Learning methods and show how to apply existing inpainting methods to remove biometric measurements as shown in Fig~\ref{fig:final}.

\section{Dataset} 
 Data for this pilot study were obtained in collaboration with the Clinical Data Warehouse of Greater Paris University Hospitals. Data were collected prospectively in Saint Louis Hospital and
 we worked closely with volunteer physicians from the NHANCE NGO to construct an heterogeneous database that includes images from various manufacturers. The database was composed of 1280 images from which 1150 were used to train and validate the method. The 130 remaining images were used to evaluate the method by a trained engineer. 
\section{Detection of the ultrasound fan area}
\label{sec:model}
We present a fully parametric method to detect the ultrasound fan area in the image and thus remove most of the annotations present outside. This is achieved by generating segmentation masks of the US region using a probabilistic approach.
\subsection{Probabilistic model of US fan area}
\begin{figure}[htb!]
\includegraphics[width=8cm]{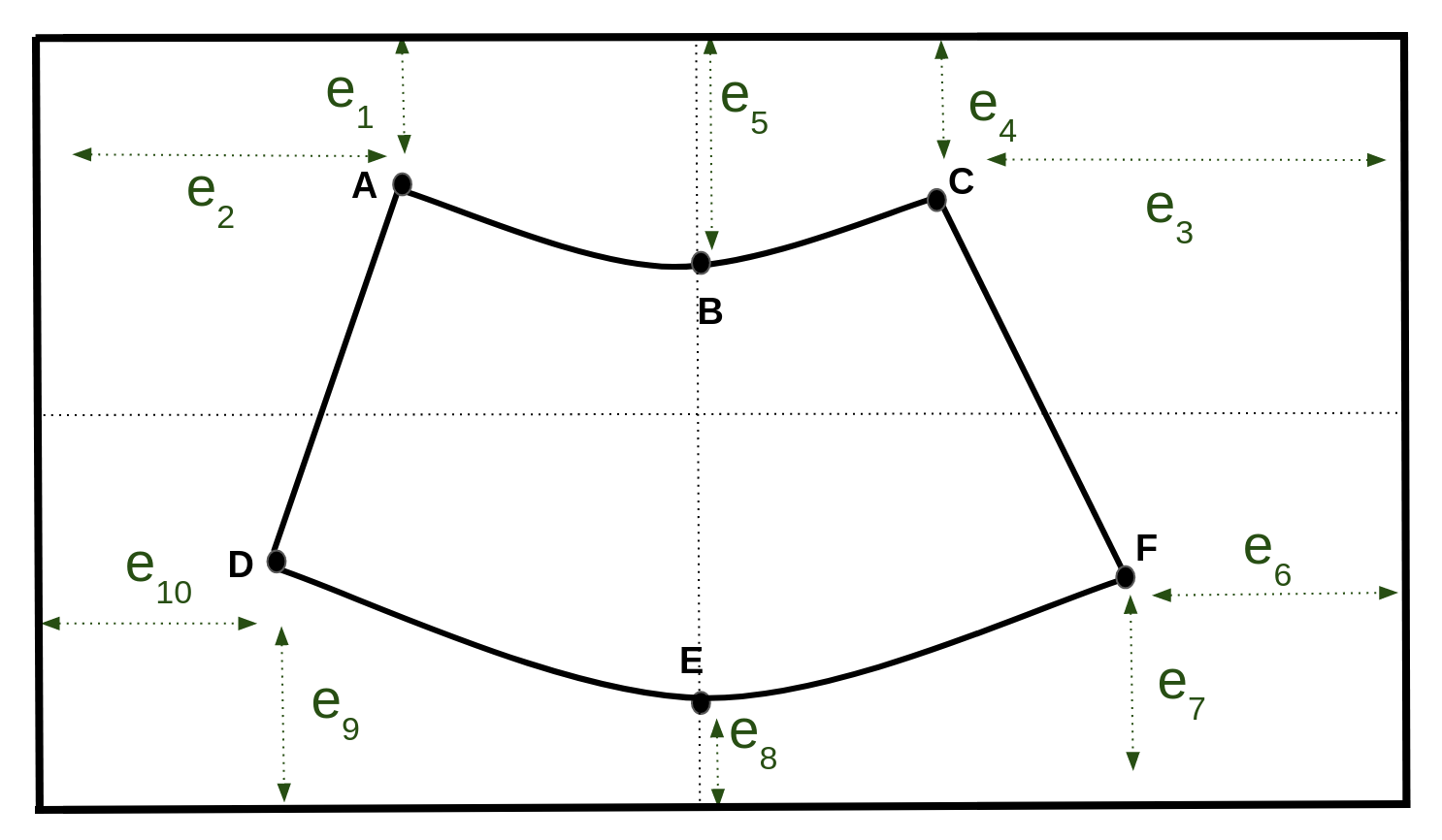}
  \caption{Parameterisation of the region of interest}
  \label{fig:eqdroite}
\end{figure}

\begin{figure}[htb!]
\begin{minipage}[b]{.48\linewidth}
  \centering
  \centerline{\includegraphics[width=4.5cm]{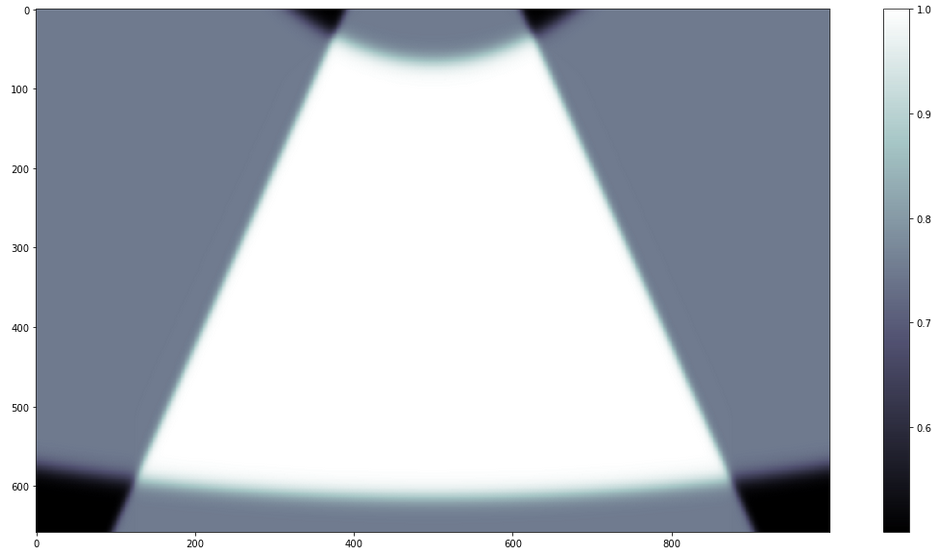}}

  \label{fig:histoback}
\end{minipage}
\hfill
\begin{minipage}[b]{0.48\linewidth}
  \centering
  \centerline{\includegraphics[width=4.5cm]{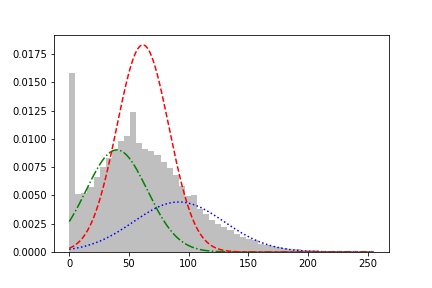}}

    \label{fig:histomixture}
\end{minipage}
\caption{(Left) Prior label probability $p(Z_i=1|\theta)$ parameterized by $\theta$; (Right) Normalized histograms of the ROI distribution. In green/blue lines the two Gaussians and in red the mixture of the Gaussians.}
\label{fig:histo}
\end{figure}

Let I be our image of size $n\cdot m$ where $n$ is the width and $m$ the height of the image. We define $\theta = \{\epsilon_{1},...,\epsilon_{10}\}$ as the set of parameters describing the truncated cone $\Omega_{\theta}$ modeling the US fan area (see Fig. \ref{fig:eqdroite}). We seek to optimize $\theta$ for every input US image via a probabilistic formulation. More precisely, we introduce the hidden binary variable  $Z_i \in [0,1]$, indicating whether voxel $i$ belongs to the fan area $\Omega_{\theta}$. The  $\theta$ parameters are equipped with a uniform prior and we propose to maximize the marginal log likelihood $p(\theta | I)$ as follows:
\begin{align*}
    \log(p(\theta |I)) &\propto \log\big(p(\theta)) + \log(p(I|\theta)\big)\\
    \log(p(I|\theta)) &= \sum_{i=0}^{n \cdot m} \log \big(p(I_{i} |Z_{i} = 1)
\cdot p(Z_{i} = 1|\theta) \\ &\quad + (p(I_{i} |Z_{i} = 0)) \cdot (1-p(Z_{i} = 1|\theta))\big) 
\end{align*}
To model the distributions, we make use of intensity values extracted from bounding boxes of all training images. The distribution of a pixel intensity in the US fan $f_i=P(I_{i} |Z_{i} = 1)$ is captured by a mixture of two Gaussians (see Fig~\ref{fig:histo}.b) whose parameters are estimated using bounding boxes of size $5\times 5$ located in the center of images. As for the background distribution $b_i=P(I_{i} |Z_{i} = 0))$ it is modeled as a uniform distribution whose parameters are estimated using bounding boxes of size $2\times2$ located in the top left, top right, bottom left and bottom right corners of the image.

$P(Z_{i} = 1|\theta)$ is the probability of voxel $i$ to be inside the fan area $\Omega_{\theta}$ which is defined in a closed form manner.  Indeed, the truncated cone region $\Omega_{\theta}$ is defined analytically as the set of points $\pos=(x, y)$ for which $f_i(\pos,\theta) \geq 0$

where $f_1,f_2$ are respectively the equations of the two straight lines $AD,CF$ and $f_3,f_4$ are respectively the equations of the two parabolas through $ABC,DEF$ as shown in Figure \ref{fig:eqdroite}. 
From this piecewise analytic description, we derive a soft implicit definition  of the fan shape by summing up the four implicit functions $f_i(\pos,\theta)$. The prior label probability is defined as Bernouilli distribution whose parameter depends on the sigmoid of the regularized implicit function~:

\begin{align*}
p(Z_{i} = 1|\theta) &= \sigma[\sum_{i=1}^4 f_i(\pos_i,\theta)]
\end{align*}
where $\pos_i$ is the position of voxel $i$ in the image.-

\subsection{Expectation-Maximization (E-M) algorithm}
Maximimizing the log joint probability is not easy since we have the log of sums. Instead, we use a lower bound which is much simpler to compute.  More precisely, we replace this maximization :
\begin{align*}
\log p(I,\theta)=
\log(p(\theta) + \log(p(I|\theta))
\end{align*}
With this one : 
\begin{align*}
    \log p(I,\theta)-D_{\mathrm{KL}}(U||p(Z|I))
\end{align*}
Where $U=\{U_n\}$ is a surrogate function for the posterior label probability $p(Z|I)$ and $D_{\mathrm{KL}}$ is the Kullback–Leibler divergence.
This can easily be solved using the E-M algorithm.
\begin{itemize}
    \item {\bf E-Step}. Compute the posterior label probability for the current value of $\theta$:
\begin{align*}
    U_i&=p(Z_i=1|I_i)\\&=\frac{r_i p(Z_i=1|\theta)}{r_i p(Z_i=1|\theta)+(1-r_i)(1-p(Z_i=1|\theta))}
\end{align*}
where \small{$r_{i}= p(I_{i} |Z_{i} = 1) \slash p(I_{i} |Z_{i} = 1)+p(I_{i} |Z_{i} = 0)$}.\\
 \item {\bf M-Step}. Find $\theta$ by maximizing the variational lower bound which is equivalent to minimizing the following expression with the current value of $U$ :
 \begin{align*}
 {\mathcal L}(\theta)&= -D_{\mathrm{KL}}(U||p(Z|I))+\log(p(\theta))\\
 &=\sum_{i=0}^{n \cdot m} -U_i\log(p(Z_i=1|\theta)) \\
 &+ (1-U_i)\log(1-p(Z_i=1|\theta)) )\\
 &+ \log(p(\theta))
 \label{eq:neglikelihood}
 \end{align*}
\end{itemize}
During the M-step we use an optimization algorithm, Limited-memory BFGS algorithm (L-BFGS). This algorithm approximates the Broyden–Fletcher–Goldfarb–Shanno algorithm (BFGS) using a limited amount of computer memory \cite{byrd1995limited,zhu1997algorithm}. 
\subsection{Reducing the computational time of the method using Deep Learning}
We use the segmentation masks generated by the method to train a neural network for the same task. This is done by training on CPU a simple U-Net on 70 \% of our dataset. We validate the method on 20 \% of our dataset and test it on the remaining 10 \% . We use the Binary Cross Entropy (BCE) with logits loss.  In inference, the processing of one frame takes approximately 0.45 seconds which is 160 times faster than the Bayesian method.

\section{Implementation and results}
\label{sec:results}
In this section we explicit the implementation details of our method and show the quantitative and qualitative results. 
\subsection{Implementation of the E-M algorithm}
\begin{figure}[htb!]
\begin{minipage}[b]{0.48\linewidth}
  \centering
  \centerline{\includegraphics[width=4.0cm]{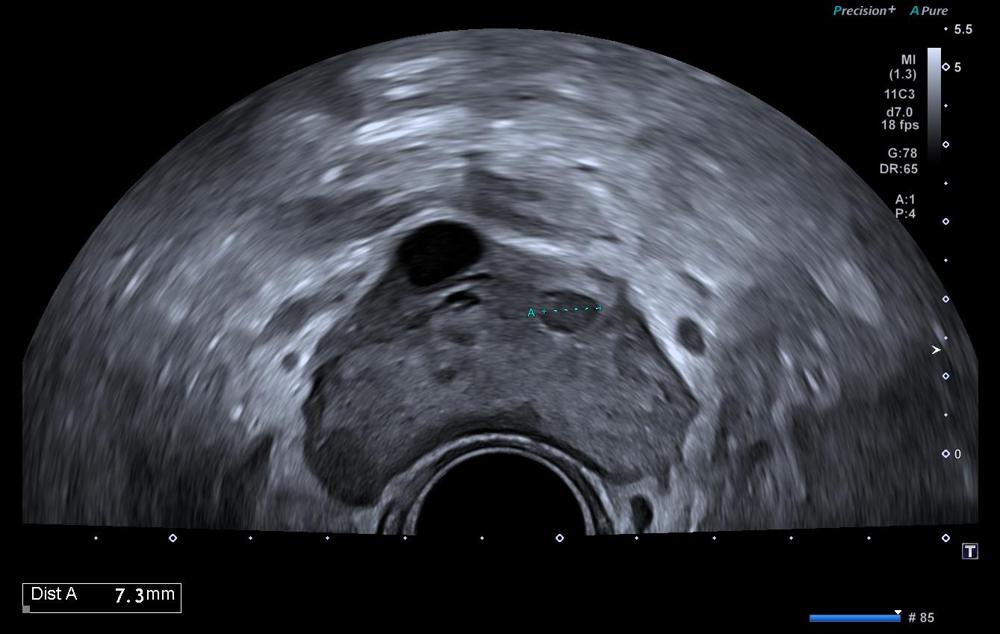}}
  \centerline{(a) Input image}\medskip
\end{minipage}
\hfill
\begin{minipage}[b]{0.48\linewidth}
  \centering
  \centerline{\includegraphics[width=4.0cm]{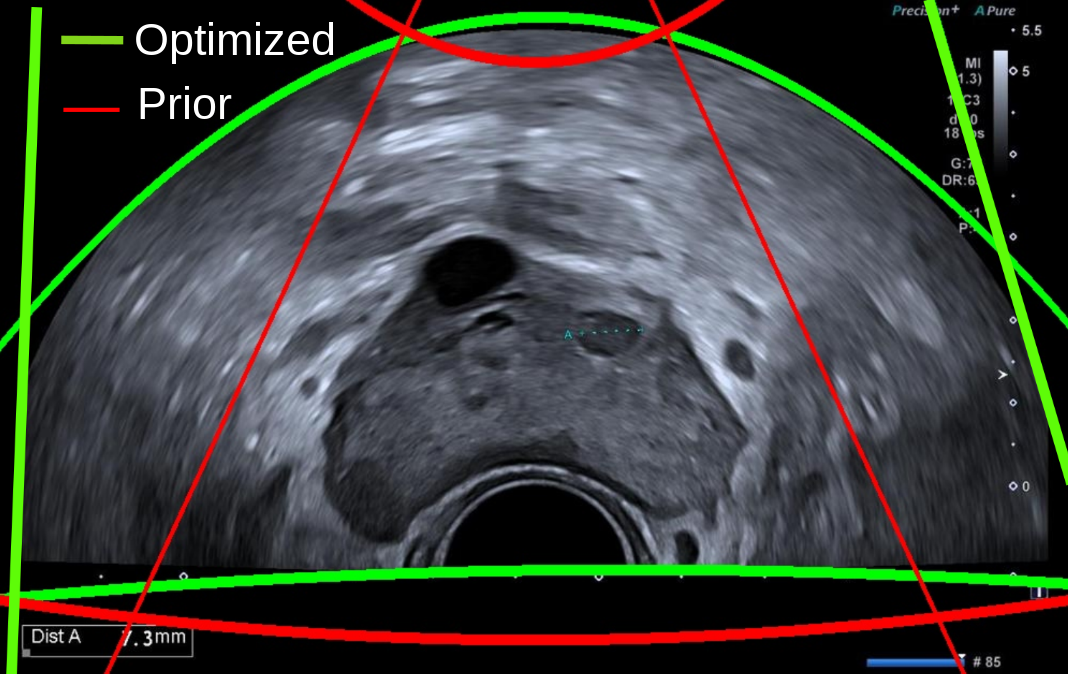}}
  \centerline{(b) ROI before and after}\medskip
\end{minipage}
\hfill
\begin{minipage}[b]{0.48\linewidth}
  \centering
  \centerline{\includegraphics[width=4.0cm]{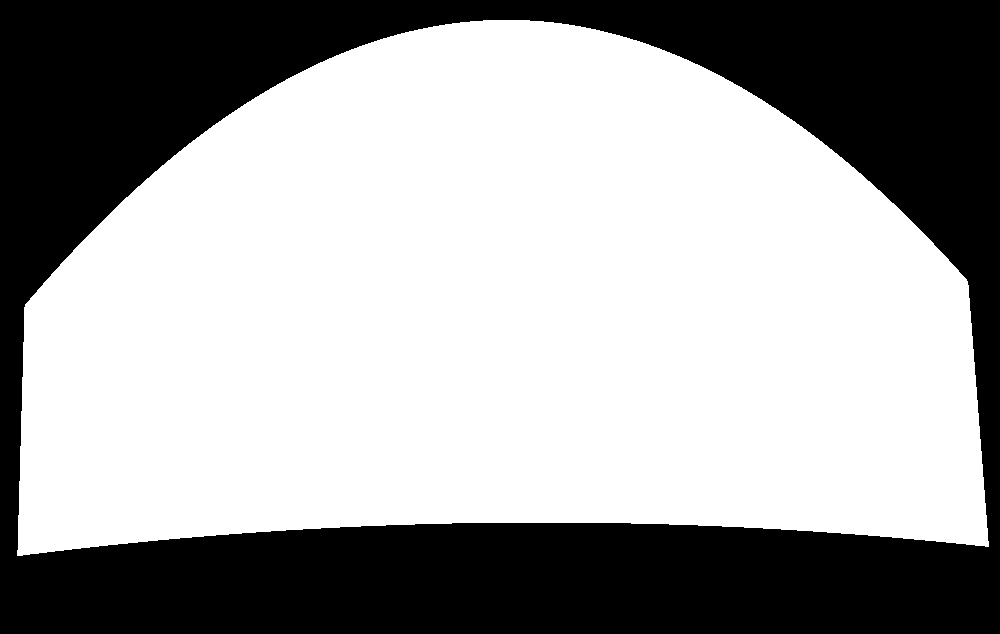}}
  \centerline{(c) Generated mask}\medskip
\end{minipage}
\hfill
\begin{minipage}[b]{0.48\linewidth}
  \centering
  \centerline{\includegraphics[width=4.0cm]{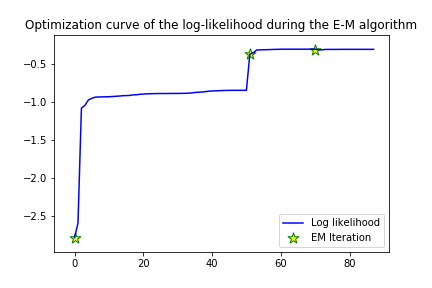}}
  \centerline{(d) Log likelihood}\medskip
\end{minipage}
\caption{Log-likelihood optimization during the EM algorithm. }
\label{fig:optim_curv}
\end{figure}

We only optimize the Kullback-Leibler term during this step since we use a uniform prior as $p(\theta)$.   
Indeed, we do not have access to a preferred range of  values for $\theta$. We show in Figure \ref{fig:optim_curv} the optimization curve of the log-likelihood during the E-M algorithm when the prior on $\theta$ is far from the ground truth. We can see that the algorithm is robust to the change of shape in the ultrasound fan area. 

\subsection{Training details of the U-Net}
We use Adam Optimizer \cite{kingma2014adam} with learning rate $0.001$ and a batch size of 1. After 10 epochs, we achieve a validation loss of 0.062 and Dice loss of 0.017. 
\subsection{Bayesian method compared to the U-Net}
\begin{figure}[htb!]
\includegraphics[width=9cm]{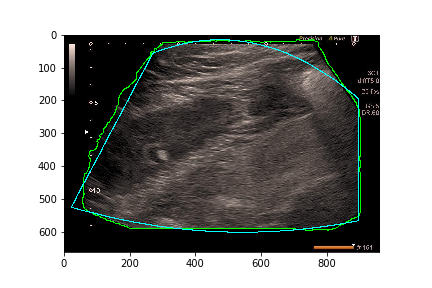}
  \caption{(Blue) Mask generated by the Bayesian method. (Green) Mask generated by the U-Net.}
  \label{fig:Compare}
\end{figure}

The Bayesian method developed in Section~\ref{sec:model} is constrained by our piecewise analytic description. It provides a good approximation of the ground truth mask, but results in a rigid delineation of the ultrasound fan area. Whereas the masks generated by the U-Net are less regularized and can therefore capture more information on the ultrasound fan area. An example of the two masks is shown in Figure~\ref{fig:Compare}. 
\subsection{Evaluation of the method}
 For the remaining 10 \% of the dataset, which we had never used before, we asked a trained engineer to evaluate the results of the method using 3 labels:
 \begin{itemize}
 \itemsep0em 
     \item Perfect match between the ultrasound fan area and the detected area.
     \item Good detection when the area of the missing part is less than $1 \%$ of the image.
     \item Poor detection when the area of the missing part is more  than $1 \%$ of the image.
 \end{itemize}

\begin{table}[t]
\centering
  \caption{Evaluation of the detection method on 130 images}
  \label{tab:table}
  \centering
  \begin{tabular}{cccc}
    \toprule
    \cmidrule{1-2}
    Label     & \# Images & Mean mismatch area   \\
    \midrule
    Perfect detection & 90 (69.2 \%) & $0.00\%$   \\
    Good detection & 37 (28.5 \%) & $0.15\%$ \\
    \midrule
    \textbf{Total} & 127 (97.7 \%) & $0.05\%$ \\
    \midrule
    Poor detection & 3 (2.3 \%)  & $5.0\%$      \\
    \bottomrule
  \end{tabular}
\end{table}

\begin{figure}[htb!]
\includegraphics[width=8.5cm]{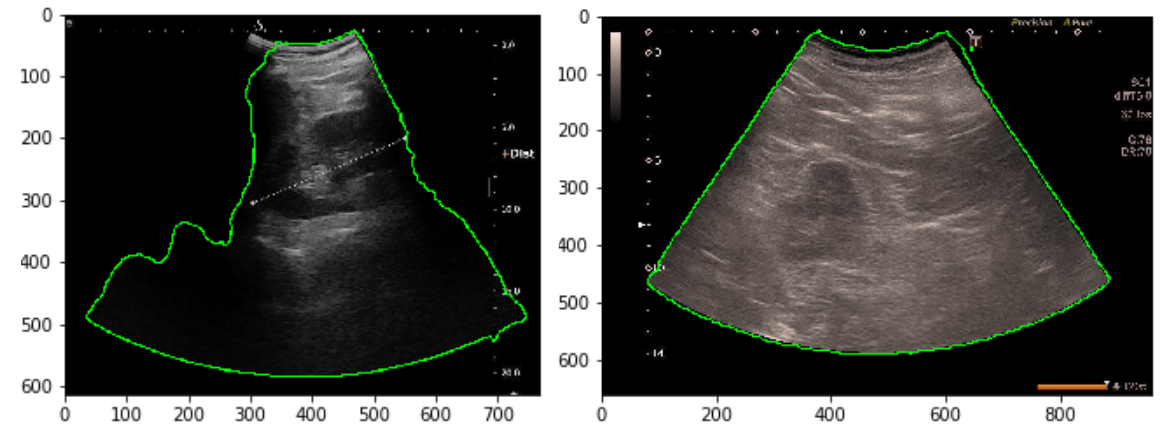}
  \caption{(Left) Example of the label 'Poor detection', a part of the fan is not detected because it is filled with low intensity pixels. (Right) Example of the label 'Good detection', the missing area corresponding to the cropped corners is barely visible to the naked eye. }
  \label{fig:detection}
\end{figure}

 We see in Table~\ref{tab:table} that 90 images were labeled as a perfect match, 37 images were labeled as good detection, with mean area of mismatch \textless 0.15 \%. This part corresponds to the corners of the fan that were slightly cropped due to the parametric definition of the fan area. Yet those tiny errors on the fan margin have no impact on the interpretation of the image content.  Finally 3 images were labeled as missing a relatively large part of the detected ultrasound fan area. This happens when a part of the ultrasound fan area is totally dark,the method then mixes up the background with the foreground. Examples are shown in Fig~\ref{fig:detection}.
\section{Inpainting images with annotations inside the ultrasound fan area}
\begin{figure}[htb!]
\setlength\abovecaptionskip{-1.0\baselineskip}
\begin{minipage}[b]{1.0\linewidth}
  \centering
  \centerline{\includegraphics[width=8.5cm]{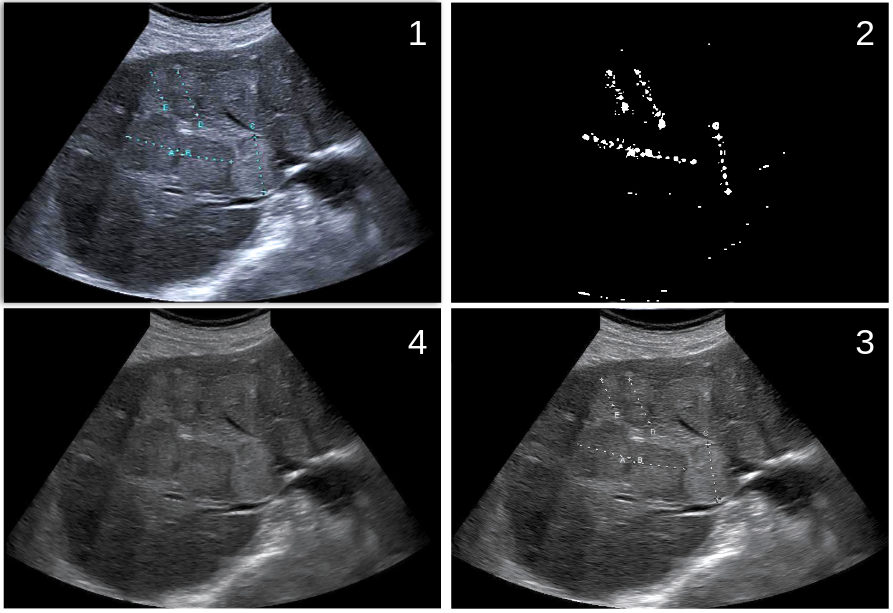}}
 \vspace{0.5cm}
\end{minipage}
\caption{Pipeline to generate masks for the inpainting algorithm. We maximize the contrast of the input image (1) and mask all pixels below a threshold value (2). We also replace all colored pixels in the image with random shades of gray (3) so that the inpainting algorithm doesn't use colored pixels present in the boundary. Finally we apply the inpaiting algorithm and denoise the resulting image using non-local-means filtering(4).}
\label{fig:mask}
\end{figure}

Here we show how to use open CV's module inpaint to replace segments and annotations present on the cone by pixels of the background. More precisely we use in-paint Telea which is based on \cite{telea2004image}. Values of pixels of the region to be inpainted are replaced by a weighted sum of neighboring pixels starting from the boundary. The challenge is to generate in a fully unsupervised way a mask of the region to be in-painted. This is done by maximizing the contrast of the image and masking all pixels below a threshold value. We also replace all colored pixels in the image with random shades of gray so that the inpainting algorithm doesn't use colored pixels present in the boundary. Finally we denoise the resulting image using non-local-means filtering \cite{buades2005non}. The method uses small patches centered on pixels. The patch of interest is compared to other patches. Then it replaces the value of the pixel by the average intensity of pixels that have patches close to the current patch. An example is shown in Fig ~\ref{fig:mask}.

\section{Conclusion}
We achieved our primary objective, namely the construction of  an automated pipeline for ultrasound fan area detection. We have shown that this method is scalable by evaluating  it on 130 varied images obtained from different machines and various shapes of the ultrasound fan area. The next step is to work closely with the Clinical Data Warehouse of Greater Paris University Hospitals, to assess the method on a larger database. The primary novelty of this method is the use of Bayesian statistics to generate training data for a Deep Learning application. We believe that this work is an important step for a larger and better exploitation of ultrasound images in the area of medical image analysis using Deep Learning. A possible improvement of the method is to replace the uninformative uniform $p(\theta)$ with a distribution estimated on the training set and include it in the M-step of the EM algorithm. The method could be further improved by adding a regularization term in the U-Net loss function. This would allow for more regular approximations of the ultrasound fan area.  We intend to release the code publicly.
 
\bibliographystyle{IEEEbib}
\bibliography{refs}

\section{Compliance with Ethical Standards}
\label{sec:ethics}

IRB approval (IRB00011591) was obtained for this multi-center study and informed consent was waived. Data for this pilot study were obtained in collaboration with the Clinical Data Warehouse of Greater Paris University Hospitals registered at the National Commission for Data Protection and Liberties (CNIL-France) under the number 1980120. 

\section{Acknowledgments}
\label{sec:acknowledgments}

This work has been supported by the French government, through the 3IA Côte d’Azur Investments in the Future project managed by the National Research Agency (ANR) with the reference number ANR-19-P3IA-0002.\\
The authors are grateful to the OPAL infrastructure from Université Côte d'Azur, the Radiology team from Saint Louis Hospital of Greater Paris University Hospitals and the French Health Data Hub for providing resources and support.\\
We thank Guillaume Oules, engineer at GoPro, for his help in the design of the NHance project. 
We thank the Clinical Data Warehouse of Greater Paris University Hospitals and especially the imaging team. 
The authors declare that they have no conflicts of interest.

\end{document}